\documentclass[11pt,a4paper]{article}            
\usepackage[english]{babel} 

\usepackage{graphicx}% Include figure files
\usepackage{epstopdf}
\usepackage{amstext,amsfonts,amsbsy,amssymb,bbm}
\newcommand {\be}[1]{\begin{eqnarray} \mbox{$\label{#1}$}  }
\newcommand{\ee}{\end{eqnarray}}

\newcommand{\pref}[1]{(\ref{#1})}

\newcommand\ie {{\it i.e.}, }

\newcommand{\nn}{\nonumber\\}

\newcommand\ra{\rightarrow }

\newcommand\half{\frac 1 2 }
\newcommand{\pdd}[2]{{\partial{#1}\over\partial{#2}}}

\renewcommand{\v}[1]{{\bf #1}}
\newcommand{\ket}[1]{|#1\rangle}
\newcommand{\bra}[1]{\langle #1 |}
\newcommand{\braket}[2]{\langle #1|#2 \rangle}
\newcommand{\com}[2]{\left[ #1, #2 \right]}

\newcommand{\cA}{ {\cal A} }

\newcommand{\ga}{ {\alpha} }
\newcommand{\gz}{ {\zeta} }
\begin{document}

\title{Explicit mapping between a 2D quantum Hall system and a 1D Luttinger liquid\\
\it{I. Luttinger parameters} }

\author{Mats Horsdal and Jon Magne Leinaas\\
Department of Physics,University of Oslo,\\ P.O.
Box 1048 Blindern, 0316 Oslo, Norway}

\date{January 14, 2006}
%\date{\today}
\maketitle
\begin{abstract} 
We study a simple model of a quantum Hall system with the electrons confined to a linear, narrow channel. The system is mapped to a 1D system which in the low-energy approximation has the form of a Luttinger liquid with different interactions between particles of equal and of opposite chiralities. We study this mapping at the microscopic level, and discuss the relation between the parameters of the 2D system and the corresponding 1D Luttinger liquid parameters. We focus in particular on how the parameters are renormalized by the electron interactions and show that the velocity parameter of the current is not modified by the interaction.
\end{abstract}

%\pacs{}

\section{Introduction}

It is a well known fact that a two-dimensional electron gas in a strong perpendicular magnetic field is dynamically equivalent to a one-dimensional electron system, when the magnetic field is sufficiently strong so that the electrons are effectively confined to the lowest Landau level. Furthermore, for the quantum Hall plateau states, where the electrons form an incompressible 2D liquid, the low energy edge excitations define a 1D chiral Luttinger liquid, as was first pointed out by X.G. Wen \cite{Wen92}.

Even if this relation between the 2D quantum Hall system and the 1D Luttinger liquid is well understood on the basis general arguments, there is no explicit mapping between these systems for general plateau states.  This leaves some uncertainty concerning the precise relation between these two systems, and we note in particular recent discussions concerning the asymptotic behaviour of the edge correlation functions of the quantum Hall system, where numerical studies have suggested a possible deviation from the expected behaviour, based on the Luttinger liquid description \cite{Goldman01,Jain01,WanEvers04}.

The motivation for the present paper is to examine in detail the relation between the quantum Hall system and the Luttinger liquid in a simplified case, where an explicit microscopic mapping between the two systems can be performed. This means that we do not consider fractional quantum Hall states,  but instead a case of integer filling. The  2D electron gas, in this model, is assumed to be constrained to a narrow channel, and the electron interaction is  assumed to be sufficiently long range to interconnect the two edges.  We examine how the  2D description of this system is mapped to a 1D theory, which in the low energy approximation is a complete Luttinger liquid theory, with both chiralities included. 

The confining potential of the 2D electrons we assume to have an harmonic form.  In the 1D description this is translated to a standard kinetic energy, which is quadratic in the particle momentum. With the electrons treated as spinless particles the model is similar to the one analyzed in the classic paper by Haldane on Luttinger liquid theory \cite{Haldane81}. Thus, we use this paper as reference for our discussion and in particular duplicate his evaluation of the Luttinger parameters by use of the bosonization technique. In these evaluations we assume a simple gaussian form for the 2D electron interaction. 

The model studied in this paper provides an explicit, microscopic description of an electron system where the appearance of the Luttinger liquid form at low energies can be studied in detail. In the standard description, an interaction which acts differently for particles with the same chirality, as compared to particles with opposite chirality, will give rise to a renormalization of the particle current \cite{Haldane81}. However,  one of our observations is that the integrated current is not modified by the  interactions. We relate this to the general argument that in the 1D system such a renormalization is not to be expected due to momentum conservation.

In the present paper our focus is on finding the precise form of  the mapping between the 2D quantum Hall system and the corresponding 1D Luttinger liquid, and on examining the renormalization effects introduced by the particle interaction. In an accompanying paper we follow up this by studying, within the same model, the electron correlation functions \cite{Horsdal}. For the 1D system this can be done by the standard bosonization technique, and the main focus of the paper is to map this to the 2D electron correlation function and to study the effects of the interaction both on the electron density and the asymptotic form of the correlation function.

%%%%%%%%%%
\section{The mapping from two to one dimensions}
%%%%%%%%%%
The possibility of a one dimensional representation of a two dimensional gas of electrons in a strong magnetic field is based on the fact that  the two orthogonal coordinates of the plane become non-commuting when projected to  the lowest Landau level. Therefore they are more like the conjugate coordinate and momentum of a 1D system than the two independent coordinates of a 2D system. The projected electron coordinates satisfy the commutation relation
\be{comrel}
\com{x}{y}= -i l_B^2
\ee 
with $l_B=\sqrt{\hbar /eB}$ as the magnetic length of the electrons in the strong magnetic field $B$.  The electron charge is $e$ and we consider $eB$ as positive. The commutator \pref{comrel} is like the commutator between  $x$ and $p$ for a particle in one dimension, except for a scaling factor which is determined by the strength of the magnetic field.
This shows that an electron wave function $\psi(x,y)$, when restricted to the lowest Landau level, is like a wave function defined on phase space rather than configuration space, and is therefore similar to a coherent state wave function. 
As a consequence of this an alternative description exists where the wave function depends on one rather than two coordinates. For example the wave functions may be of the form $\psi(x)$, with $x$ as the one-dimensional coordinate and $y$ represented by a momentum operator $y=il_B^2\pdd{}{x}$. 

Clearly the mapping from 2D to 1D is not unique. We may use $y$ rather than $x$ as the new coordinate, but we may also map the system to a circle rather than an infinite line. Thus, by using the the polar coordinates $\phi$ and $r^2$ as conjugate variables, the 1D wave function may be chosen as a function of the angle variable, $\psi(\phi)$. In this case $r^2$ represents the angular momentum operator,  $r^2=-2il_B^2\pdd{}{\phi}$, with the restriction on the angular momenta to be non-negative. However, in the present case, with the electrons confined to a linear channel, where $x$ is chosen as the coordinate along the channel, the natural choice is to use $x$ as the coordinate in the one-dimensional description.

An explicit mapping from 2D to 1D can readily be made if we make the proper gauge choice. We focus first on the non-interacting case, with no confining potential, and with the electrons here as in the rest of the paper treated as spinless particles. We choose the Landau gauge, with the vector potential expressed in terms of the constant perpendicular magnetic field as
\be{Landaugauge}
A_x=-yB\;,\quad A_y=0
\ee   
In this gauge the single particle Hamiltonian is
\be{Ham1}
H&=&{1\over {2m}}(\bf p-e\bf A)^2\nn
&=&{1\over {2m}}(p_x+eBy)^2+{1\over {2m}}p_y^2\nn
&=&{1\over {2m}}(p_y^2+m^2\omega_c^2(y+{l_B^2\over \hbar}p_x)^2))
\ee
where in the last expression the cyclotron frequency $\omega_c=eB/m$ has been introduced. The Hamiltonian is explicitly translationary invariant in the x-direction, and therefore energy eigenstates that are also eigenstates of $p_x$ can be defined. With the momentum quantized as $p_x=\hbar k$, the Hamiltonian is reduced to the form of a 1D harmonic oscillator, with potential minimum at the $k$-dependent position 
\be{yk}
y_k= -{1 \over eB}\hbar k=-l_B^2 k
\ee
The corresponding ground state eigenstates have the form
\be{pxeigen}
\psi_k(x,y)={\cal N} e^{i kx}\psi_0(y-y_k)
\ee
with  $\cal N$ as a normalization constant and $\psi_0(y-y_k)$ as the ground state of the (effective) harmonic oscillator equation in the $y$-direction,
\be{harmosc}
\psi_0(y-y_k)=(\frac{1}{\pi l_B^2})^{\frac{1}{4}}e^{-\frac{1}{2 l_B^2}(y-y_k)^2}
\ee
These states define the lowest Landau level, with complete degeneracy in energy due to independence of the value of $k$.

In the following we restrict $x$ to be a periodic variable of period $L$, which means that the 2D system is restricted to a cylinder of circumference $L$. The normalization is then ${\cal N}=\sqrt{1/L}$ and assuming periodic boundary conditions the momentum $k$ is restricted to $k=2\pi n/L$, with $n$ as an integer. Since $L$ is considered as a regularization parameter, much larger than other relevant length scales of the system, we will take the limit $L\to \infty $ at any time it is convenient in order to simplify expressions.

A general state restricted to the lowest Landau level has the form
\be{genfunk}
\psi(x,y)=\sum_k c_k\psi_k(x,y) = {1\over\sqrt{L}}\sum_k c_k\;e^{ikx}\psi_0(y-y_k)
\ee
Since the $y$-dependence in this representation is fixed by the harmonic oscillator wave function, the state can be specified by the the momentum coefficients $c_k$ alone. Thus, the state is fully described by a 1D wave function defined as
\be{genfunk2}
\psi(\xi)=\sum_k c_k\psi_k(\xi) = {1\over\sqrt{L}}\sum_k c_k \;e^{ik\xi}
\ee
where $\xi$ is  a periodic variable in 1D, with the same period $L$ as $x$. Even if $\xi$ is essentially identical to the $x$-coordinate, it is convenient to make a distinction between this 1D coordinate and the corresponding  2D coordinate $x$.

The form \pref{genfunk} of the wave function explicitly shows the one-dimensionality of the lowest Landau level, with a simple one-to-one mapping between the 2D and 1D representations conveyed by the expansion coefficients $c_k$. In the Dirac bra-ket formalism we write the wave functions as
\be{braket}
\psi(x,y)&=&\braket{x,y}{\psi}\nn
\psi(\xi)&=&\braket{\xi}{\psi}
\ee
with $\ket{x,y}$ and $\ket{\xi}$ corresponding to two different, complete sets of basis states in the lowest Landau level. An explicit expression for the transition between the 2D and 1D wave functions is given by
\be{transfunc}
f(x-\xi,y)\equiv\braket{x,y}{\xi}={1\over L}\sum_k e^{ik(x-\xi)}\psi_0(y-y_k)
\ee
The transition function only depends on the relative x-coordinate $x-\xi$, and by introducing the expression for the harmonic oscillator wave function the transition function can be written as
\be{transfunc2}
f(x,y)={1\over {L\sqrt{l_B\sqrt{\pi}}}}\sum_k \exp\left[ikx-{1\over{2l_B^2}}(y+l_B^2k)^2\right]
\ee
In the limit $L\ra\infty$ an expression in closed form can be found
\be{transfunc3}
f(x,y)&=&{1\over {2\pi\sqrt{l_B\sqrt{\pi}}}}\int_{-\infty}^{\infty} dk \exp\left[ikx-{1\over{2l_B^2}}(y+l_B^2k)^2\right]\nn
&=&{1\over {\sqrt{2\pi\sqrt{\pi}\,l_B^3}} }\exp\left[-{1\over{2l_B^2}}x(x+2iy)\right]
\ee
We note that in the $x$-direction the transition function is exponentially damped with the magnetic length as the damping length. This means that, although the mapping from 2D to 1D is nonlocal, the transformation essentially restrict the variables $x$ and $\xi$ to be equal within a magnetic length.

The inverse transformation is given by the complex conjugate function
\be{contrans}
f^*(x-\xi,y)=\braket{\xi}{x,y}
\ee
and since the transition function has the symmetries
\be{transsym}
f^*(x,y)=f(x,-y)=f(-x,y)
\ee
the mapping between the 1D and 2D wave functions can be expressed as
\be{transwave}
\psi(\xi)&=&\int dx dy f(\xi-x,y)\psi(x,y)\nn
\psi(x,y)&=&\int d\xi f(x-\xi,y)\psi(\xi)
\ee

%%%%%%%%%
\section{Mapping operators from two to one dimension}
%%%%%%%%%%
The mapping \pref{transwave} can in a straightforward way be extended to a transformation between operators in the 2D and 1D descriptions. Thus, the matrix element of an arbitrarily chosen operator can be written
\be{Amatrix}
\bra{\psi}A\ket{\phi}&=&\int d^2x \int d^2 x' \psi^*(\v r) A(\v r,\v r') \phi(\v r') \nn
&=&\int d\xi \int d\xi' \psi^*(\xi) A(\xi,\xi')\phi(\xi')
\ee
where the first line refers to the 2D formulation and the second line to the 1D formulation. (In the first expression the vector notation $\v r=(x,y)$ has been used for the 2D coordinates.) From the transformation formula for the wave functions, this gives the following relation between the integration kernels of the operator in two and one dimension,
\be{transop}
A(\xi,\xi')=\int d^2 x \int d^2 x' f(\xi-x, y) f^*(\xi'-x',y') A(\v r, \v r')
\ee
with inverse
\be{transop2}
A(\v r,\v r')=\int d\xi \int d\xi' f^*(\xi-x, y) f(\xi'-x',y') A(\xi, \xi')
\ee

A potential that is local in 2D has the form
\be{locop}
V(\v r,\v r')=V(\v r)\delta(\v r-\v r')
\ee
and this gives an operator that is non-local in 1D
\be{nonloc}
V(\xi,\xi')&=&\int d^2 x  f(\xi-x, y) f^*(\xi'-x,y) V(x,y)\nn
&=&{1\over {2\pi\sqrt{\pi}\,l_B^3}}\int dx dy\exp\left[-{1\over{2l_B^2}}\{(x-\xi)^2+(x-\xi')^2+2i(\xi-\xi')y)\}\right]V(x,y)\nn
\ee
The non-locality is now determined partly by the  magnetic length, but also in part by the range of the potential.

For a two-particle operator $A$ similar expressions are found by relating the 2D matrix element
 \be{twopart2}
 A(\v r_1,\v r_2;\v r'_1,\v r'_2)=\bra{\v r_1,\v r_2}A\ket{\v r'_1,\v r'_2}\nn
  \ee
  to the 1D matrix element
   \be{twopart}
  A(\xi_1,\xi_2;\xi'_1,\xi'_2)=\bra{\xi_1,\xi_2}A\ket{\xi'_1,\xi'_2}
  \ee
We restrict the operator to be be a two-particle potential V, which depends only on the relative coordinates in a local way, and which is antisymmetric in the particle coordinates,
\be{intpot}
 V(\v r_1,\v r_2;\v r'_1,\v r'_2)= V(\v r_1-\v r_2)(\delta(\v r_1-\v r'_1)\delta(\v r_2-\v r'_2)-\delta(\v r_1-\v r'_2)\delta(\v r_2-\v r'_1))
 \ee
The corresponding 1D operator will then also depend only on the relative coordinates, but now in a non-local way,
 \be{1dinter}
   V(\xi_1,\xi_2;\xi'_1,\xi'_2)=  V(\xi_1-\xi_2;\xi'_1-\xi'_2)\delta(\half(\xi_1+\xi_2)-\half(\xi'_1+\xi'_2))
   \ee
Written as functions of the relative coordinates the correspondence between the 1D and 2D potentials is of the form
  \be{1dinter2}
   V(\xi;\xi')=\int d^2 x \;g(\xi,\xi';x,y) V(x,y)
   \ee
where the two-particle $g$-function, which is determined from the one-particle $f$-function, has the form
\be{gtrans3}
g(\xi,\xi';x,y)={1\over {(\sqrt{2\pi}\,l_B)^3}}{\cal A} \exp\left[-{1\over{4l_B^2}}\{(x-\xi)^2+(x-\xi')^2+2i(\xi-\xi')y)\}\right]
\ee
In this expression $\cA$ denotes antisymmetrization with respect to interchange of the particle coordinates, $\xi\to-\xi$ and $\xi'\to-\xi'$.

A gaussian interaction, which we shall apply in the following,
\be{gaussint}
V(\v r)=V_0e^{-\ga^2 \v r^2}
\ee
gives rise to the following two-particle interaction in 1D 
\be{1dgauss}
V(\xi,\xi') = \frac{V_0}{2\sqrt{\pi(1+2\ga^2l_B^2)}\ga l_B^2}\cA \exp{[-\frac{\ga^2}{4(1+2\ga^2l_B^2)}(\xi+\xi')^2-\frac{(1+2\ga^2l_B^2)}{16 l_B^4\ga^2}(\xi-\xi')^2]}\nn
\ee
which for a long range potential, $\ga l_B<<1$, can be approximated by
\be{1dgauss2}
V(\xi,\xi') = \frac{V_0}{\sqrt{\pi}\ga l_B^2}\cA \exp{[-\frac{\ga^2}{4}(\xi+\xi')^2-\frac{1}{16 l_B^4\ga^2}(\xi-\xi')^2]}
\ee

We note that the 1D interaction given by \pref{1dgauss} and \pref{1dgauss2} involves two length scales. The first one is given by the  length  $1/\ga$, which is the range of the interaction potential in 2D, while the other is the shorter length $\ga l_B^2$. The appearance of two scales rather than one comes from the difference in how the interaction in the x-direction and the y-direction in 2D are treated by the mapping to 1D. As a result of this the interaction in 1D is not a local density interaction. We may interpret the largest length $1/\ga$ as determining the range of the interaction, while the shorter length $\ga l_B^2$ determines the degree of non-locality.

In $k$-space the interaction has the following form
\be{kspace}
V(k,k')&=&\int d\xi \int d\xi' \;V(\xi,\xi') \exp(-i(k\xi-k'\xi'))\nn
&=& \bar V_0\; \cA \exp[-\frac{1}{4 a^2}(k-k')^2-b^2(k+k')^2]
\ee
where we have introduced the following constants
\be{newconst}
\bar V_0&=&2\sqrt{\frac{\pi}{1+2\ga^2l_B^2}}\frac{V_0}{\ga}\approx \frac{2\sqrt{\pi}V_0}{\ga}\nn
a&=&\frac{\ga}{\sqrt{1+2\ga^2l_B^2}}\approx \ga \nn
b&=&\frac{\ga l_B^2}{\sqrt{1+2\ga^2 l_B^2}} \approx \ga l_B^2
\ee
The approximations indicated in the expressions above are based on the assumption  $\ga l_B<<1$. 
When the confining potential is introduced in the next section, the form of the interaction \pref{kspace} is still valid, but with $\bar V_0$, $a$ and $b$ modified by the potential. We shall therefore use this form in the further discussion of the interaction. Furthermore, we shall throughout the paper assume the interaction to be repulsive, so that $\bar V_0>0$.

%%%%%%%%%
\section{Introducing the confining potential}
%%%%%%%%%%
When the confining potential is introduced, the degeneracy of the lowest Landau level is lifted. We assume the potential to have a harmonic form in the y-direction, and the single particle Hamiltonian then is
\be{Ham2}
H&=&{1\over {2m}}(p_x+eBy)^2+{1\over {2m}}p_y^2+\half m\omega^2y^2\nn
&=&{1\over {2m}}p_y^2+\half m[(\omega_c y+{{p_x}\over{m}})^2+\omega^2y^2]
\ee
with $\omega$ as the oscillator frequency. 
It is well known that such a harmonic oscillator potential  can be absorbed in a stronger effective cyclotron frequency $\bar\omega_c$ in the following way,
\be{Ham3}
H={1\over {2m}}p_y^2+\half m\bar{\omega}_c^2(y+\frac{\omega_c}{m\bar{\omega}_c^2}p_x)^2 +{1\over {2m}}\frac{\omega^2}{\bar{\omega}_c^2}p_x^2
\ee
with
\be{effcycl}
\bar{\omega}_c=\sqrt{\omega_c^2+\omega^2}
\ee

The first two terms of \pref{Ham3} can be identified as the Hamiltonian of an electron in an effective magnetic field $\bar B=\sqrt{B^2+m^2\omega^2/e^2}$. Since the $x$-momentum is a constant of motion, it can be quantized, $p_x=\hbar k$,  and the energy eigenstates therefore have the same form as before, although with a modified magnetic length, $\bar l_B=\sqrt{\hbar/e\bar B}$. Note however a small modification of the $k$-dependent shift in the minimum of the harmonic oscillator potential in the $y$-direction,
$y_k=-(\omega_c/\bar\omega_c)\bar l_B^2 k$. 

A modified lowest Landau level  can now be defined by reference to the lowest energy states of the electron in the effective magnetic field $\bar B$. A further mapping of 1D can be done as previously discussed, but now with parameters referring to the effective magnetic field rather than the true magnetic field. The precise form for the f-function which defines the mapping between wave functions in 2D and 1D is
\be{transfunc4}
f(x,y)&=&{1\over {2\pi\sqrt{\bar l_B\sqrt{\pi}}}}\int_{-\infty}^{\infty} dk \exp\left[ikx-{1\over{2\bar l_B^2}}(y+{{\bar l_B^4}\over{l_B^2}}k)^2\right]\nn
&=&{{\Lambda}\over{\sqrt{2\pi\sqrt{\pi}\;{\bar{l}_B}^3}}}\;\exp\left[-{1\over{2\bar l_B^2}}\Lambda x(\Lambda x+2iy)\right]
\ee
with 
\be{Lambda}
\Lambda=\left(\frac{l_B}{\bar l_B}\right)^2=\frac{\bar{\omega}_c}{\omega_c}=\sqrt{1+\frac{\omega^2}{\omega_c^2}}
\ee
We note that the transformation now does not only depend on the effective magnetic length, but also the dimensionless parameter $\Lambda$.

The rescaling of the magnetic length, and the effective rescaling of the $x$-coordinate in the expression for $f(x,y)$ implies a similar rescaling of the parameters in the expressions for the 1D interaction. Thus, the form of the k-space interaction \pref{kspace} is unchanged, but with rescaled parameters,
\be{rescaled}
\bar V_0&=&2\sqrt{\frac{\pi}{1+2\ga^2\bar l_B^2}}\;\frac{V_0}{\ga}\approx \frac{2\sqrt{\pi}V_0}{\ga}\nn
a&=&\Lambda\frac{\ga}{\sqrt{\Lambda^2+2\ga^2\bar l_B^2}}\approx\Lambda\ga \nn
b&=&\frac{\ga\bar  l_B^2}{\Lambda\sqrt{1+2\ga^2\bar  l_B^2}} \approx\frac{\ga} { \Lambda} {\bar l}_B^2
\ee

The states of the lowest Landau level are no longer degenerate in energy due to the last term of \pref{Ham3}. Thus, the $k$ dependent single-particle energies are $E_k=(\hbar^2 k^2/2m)(\omega/\bar\omega_c)^2+\hbar\bar\omega_c/2$. Clearly, when $k$ is sufficiently large the energy will exceed the excitation energy to the next Landau level. Therefore the ground state of the $N$-particle system for sufficiently large $N$ will include occupied single particle states also in higher Landau levels. In our case we assume the confining potential to be sufficiently soft so that this is avoided. To be more precise, let us consider $N_0=k_F L/\pi$ as the relevant particle number, with the ground state corresponding to occupation of all states in the interval $-k_F<k<k_F$ between the two Fermi points.  With $W$ denoting the physical width of the corresponding band of states in the 2D plane, we have $W=2(\bar l_B^4/l_B^2)k_F$ and the condition for the states of highest energy, at the edges of this band, to have lower energy than the excitation energy to the next Landau level is
\be{soft}
\frac{\omega}{\omega_c} \lesssim\;\frac{\bar l_B}{W}<<1
\ee
The last inequality follows from the assumption that even if the channel of 2D electrons is narrow, so that interactions between the two edges is possible, it is wide at the  length scale set by the magnetic field.

Since the oscillator frequency is much smaller than the cyclotron frequency, the difference between $\omega_c$ and $\bar\omega_c$ (and between $l_B$ and $\bar l_B$) is small and can in most cases be neglected. This means for the mapping between the 2D and 1D descriptions that $\Lambda\approx 1$ and $\bar l_B\approx l_B$ and therefore the original form of the transformation functions, introduced in Sects. 2 and 3, can be used to a good approximation,  also in the presence of the confining potential.

%%%%%%%%%%
\begin{figure}[h]
\label{paths}
\begin{center}
\includegraphics[height=3.5cm]{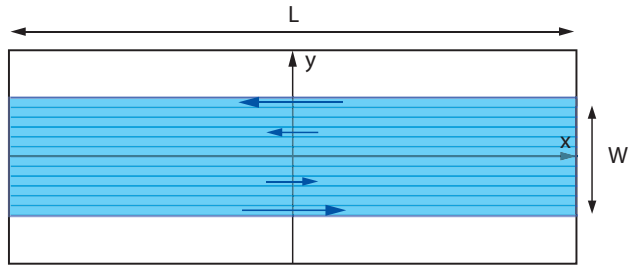}
\end{center}
\caption{\small The occupied 2D electron states of the many-electron ground state. Due to the harmonic oscillator potential in the y-direction, the electrons are confined to a band of width $W$ (indicated by blue color). The electron states are localized in the y-direction and are extended in the x-direction, with a periodic length $L$. The potential gives rise to a current density in the x-direction that changes linearly with $y$. }
\end{figure}
%%%%%%%%%%

When interactions between the electrons are introduced in the next section this picture does not change as long as these are sufficiently weak. However, for a stronger interaction it is natural to consider the confining potential discussed here as being the effective potential defined by an attractive background potential together with  the (repulsive) potential produced by the electrons in the Fermi sea. In that case the energy constraints given above should be interpreted as applying to this effective potential.

A physical effect of the confining potential is to introduce drift currents along the quantum Hall channel, with opposite directions of the current on the two sides of the channel. For the harmonic oscillator potential the currents are not confined to the edges, but are distributed in a continuous way through the many-electron system, with maximum values at the edges. However, the edges themselves are well defined, since the electron density in the ground state is essentially constant through the sample and falls abruptly to zero at the edges, within a width of one magnetic length.

%%%%%%%%%%%%%%%
\section{ The interacting system and its low-energy approximation}
%%%%%%%%%%%%%%%
In the 1D representation the $N$-particle Hamiltonian  has the form
\be{Ham1D}
H_N=\sum_{i=1}^N{{1}\over{2M}}p_i^2+{{N}\over{2}}\hbar \bar{\omega}_c+\sum_{i<j}V_{ij}
\ee
where $V_{ij}$ is the non-local two-body interaction, with matrix elements previously discussed in Sect.~3, and $M$ is the effective 1D mass, related to the electron mass by
\be{effmass}
M=m\,\frac{\bar{\omega}_c^2}{\omega^2}
\ee
In the second quantized form the Hamiltonian is
\be{secondHam}
H=\sum_k({{\hbar^2}\over{2M}}k^2+\half\hbar\bar{\omega}_c)c_k^{\dag}c_k+\frac{1}{4L}\sum_{q,k_1,k_2}V(k,k')c_{k_1}^{\dag}c_{k_2}^{\dag}c_{k_2-q}c_{k_1+q}
\ee
with $k=(k_1-k_2)/2$ and  $k'=(k_1-k_2)/2+q$. $V(k,k')$ is given by \pref{kspace}, but without the antisymmetrization $\cA$ in the variables $k$ and $k'$. The explicit antisymmetrization is not needed due to the anticommutation relations between the $c$ operators, and the chosen unsymmetrized form is convenient when discussing the low energy approximation.  

We focus now on the low energy approximation to the Hamiltonian, which means that we restrict the system to include only low energy excitations around the ground state of a system for a given number $N_0$ of electrons. 
These excitations are in a natural way characterized by their chirality, where the excitations close to the Fermi point $+k_F$ have positive chirality and correspond to excitations of rightmoving electrons, whereas excitations close to $-k_F$ have negative chirality, corresponding to excitation  of leftmoving electrons. We characterize these two possibilities by the chirality parameter $\chi=\pm 1$.

The low energy approximation is most conveniently introduced by rewriting the Hamiltonian in terms of operators that are normal ordered relative to the filled Fermi sea. Thus, the first part of the Hamiltonian is written as
\be{H0}
H_0= \sum_k({{\hbar^2}\over{2M}}k^2+\half\hbar\bar{\omega}_c):c_k^{\dag}c_k:
+ \sum_{k=-k_F}^{k_F}({{\hbar^2}\over{2M}}k^2+\half\hbar\bar{\omega}_c)
\ee
and since the normal ordered operator, in the low energy approximation, can be regarded as giving non-vanishing contributions only when $k$ is close to $k_F$ or $-k_F$, a linearization of the $k$-dependent function around these points can be performed. With the contributions from the two Fermi points identified by their different values of the chirality parameter $\chi$, the kinetic term $H_0$ can then be written as
\be{}
H_0&=& v_F\hbar \sum_{\chi ,\,k}(\chi k-k_F):c_{\chi,\, k}^{\dag}\,c_{\chi,\, k}:
+ (\half M v_F^2+\half \hbar \bar\omega_c)(N-N_0)\nn
&&+\sum_{k=-k_F}^{k_F}({{\hbar^2}\over{2M}}k^2+\half\hbar\bar{\omega}_c)
\ee
with $v_F=\hbar k_F/M$ as the Fermi velocity. Even if in the first term the summation variable $k$ is initially restricted to be close to $\chi k_F$, we follow the usual approach to treat $\chi$ and $k$ as independent variables, since in the low energy approximation there is effectively no contribution to the normal ordered operators from the additional values of $k$.

The interaction part of $H$ is treated in a similar way,
\be{inter}
V&=&\frac{1}{4L}\sum_{q,k_1,k_2}V(\half(k_1-k_2),\half(k_1-k_2)+q)):c_{k_1}^{\dag}c_{k_1+q}:\;:c_{k_2}^{\dag}c_{k_2-q}: \nn
&&+\frac{1}{2L} \sum_{k}\left[\sum_{k'=-k_F}^{k_F}V(\half(k-k'),\half(k-k'))\right]:c_{k}^{\dag} c_{k}:\nn
&&-\frac{1}{4L}\sum_{k} V(\half k,\half k)\;N\nn
&&+\frac{1}{4L}\sum_{k_1=-k_F}^{k_F}\sum_{k_2=-k_F}^{k_F}V(\half(k_1-k_2),\half(k_1-k_2))
\ee
We first consider the first term of \pref{inter}, which is the two-particle interaction. In the low energy approximation $k_1$ and $k_2$ are restricted to be close to one of the Fermi points, and we make a distinction between two contributions from this term, with the first contribution corresponding to interaction between particles with the same chirality and the other to interaction between particles with opposite chiralities. 

For the case of equal chiralities we have
\be{same}
V(\half(k_1-k_2),\half(k_1-k_2)+q))&=&\bar V_0\; \exp[-\frac{q^2}{4 a^2}-b^2(k_1-k_2+q)^2]\nn
&\approx&\bar V_0\; \exp[-\frac{q^2}{4 a^2}]\nn
&\equiv& V_1(q)
\ee
where we have neglected the contribution from the $b$ term due to the smallness of $b$. Thus, if the range of the potential is comparable to the width $W$ of the quantum Hall sample we have  $b\approx 1/k_F$ , and since in this case $|k_1-k_2+q|<<k_F$ the contribution is negligible.
However, when the two particles are at opposite edges this is no longer the case, since we then have $|k_1-k_2+q|\approx 2k_F$. In that case the following approximation is valid
\be{opposite}
V(\half(k_1-k_2),\half(k_1-k_2)+q))&=&\bar V_0\; \exp[-\frac{q^2}{4 a^2}-b^2(k_1-k_2+q)^2]\nn&\approx&\bar V_0\;\exp[-4b^2 k_F^2] \exp[-\frac{q^2}{4 a^2}]\nn
&\equiv& V_2(q)
\ee

Since, in both cases, the momentum transfer $q$ is effectively restricted by $|q|<<k_F$, due to the exponential damping term $\exp(-q^2/4a^2)$, the first term of the interaction \pref{inter} can be approximated by
\be{2part}
V_{\rm 2part}=\frac{1}{4L} \sum_{\chi,\,q}(V_1(q)\rho_{\chi,\,q}\rho_{\chi,\,-q}+V_2(q)\rho_{\chi,\,q}\rho_{-\chi,\,-q})
\ee
where we have introduced the chiral density operators
\be{density}
\rho_{\chi,\, q}=\sum_k :c_{\chi,\,k}^{\dag}c_{\chi,\,k+q}:
\ee
Although the original interaction is non-local, we note that the two contributions to the low-energy interaction, $V_1$ and $V_2$, are both local interactions.  The original non-locality has given rise to a difference between these terms, in the form of a constant, $b$-dependent factor,
\be{V1V2}
V_2(q)=  \exp[-4b^2 k_F^2]V_1(q)
\ee
 
We next consider the second term of the interaction \pref{inter}, which is a one-particle operator that can be treated in the similar way as $H_0$. We write this term as
\be{1part}
V_{\rm 1part}=\frac{1}{2L}\sum_k U(k) :c_k^{\dag}c_k:
\ee
with
\be{U}
U(k)=\sum_{k'=-k_F}^{k_F} V(\half(k-k'),\half(k-k'))
\ee
In the low energy approximation $k$ has to be close to either $k_F$ or $-k_F$. We separate these two contributions, corresponding to the two values of the chirality parameter $\chi$, and write $k=\chi k_F+(k-\chi k_F)$ with $|k-\chi k_F|<<k_F$. By expanding to first order in the small quantity $k-\chi k_F$ we get the following two low energy contributions
\be{Uchi}
U_{\chi}(k)=\frac{L}{2\pi}\bar V_0\int_0^{2k_F}e^{-b^2k^2}dk-\frac{L}{2\pi}\bar V_0(1-e^{-4b^2k_F^2})(\chi k-k_F)
\ee
where only the second term is $k$-dependent. In this approximation the one-particle operator is
\be{1part2}
V_{\rm 1part}=-\frac{\bar V_0 }{4\pi}(1-e^{-4b^2k_F^2})\sum_{\chi ,\,k}(\chi k-k_F) :c_{\chi,\,k}^{\dag}c_{\chi,\,k}:
+\frac{\bar V_0 }{4\pi}[\int_0^{2k_F}e^{-b^2k^2}dk]\; (N-N_0)\nn
\ee
where the second term is proportional to the particle number $N-N_0$. The first term is linear in the momentum variable $k$ and effectively modifies the Fermi velocity by an interaction dependent contribution,
\be{Fermi}
\bar v_F=v_F-\frac{\bar V_0 }{4\pi\hbar}(1-e^{-4b^2k_F^2})
\ee
In the 2D description this correction to the Fermi velocity can be interpreted as due to a correction to the confining potential caused by interactions with electrons in the bulk of the system.

The full Hamiltonian in the low-energy approximation can now be found by collecting the terms discussed above. We modify further the expression by adding a term proportional to $N$ (interpreted as an effective chemical potential) and by removing the constant ground state energy. The resulting expression for the Hamiltonian is
\be{fullHam}
H=\bar v_F\hbar \sum_{\chi ,\,k}(\chi k-k_F):c_{\chi,\, k}^{\dag}\,c_{\chi,\, k}:+\frac{1}{4L} \sum_{\chi,\,q}(V_1(q)\rho_{\chi,\,q}\rho_{\chi,\,-q}+V_2(q)\rho_{\chi,\,q}\rho_{-\chi,\,-q})\nn
\ee
This is precisely of the Luttinger form, as discussed in ref.  \cite{Haldane81}, and the Hamiltonian can therefore be diagonalized by the same bosonization technique. Note, however, that $\bar v_F$ already include contributions from the interactions. Such contributions are not included in \cite{Haldane81}, where the interactions are introduced directly in the normal ordered form \pref{fullHam}. Also note that our derivation of the interaction term from the underlying 2D model has given a  difference in the strength of $V_1$ and $V_2$, as a consequence of  the non-locality that is introduced by the mapping from 2D to 1D. In a truly 1D theory  such a difference may seem somewhat unnatural, since it does not correspond to a local density interaction. 

Although the original Hamiltonian \pref{secondHam} in principle include chirality changing terms, these are (due to momentum conservation) suppressed in the low energy approximation. Thus, the chirality charge
\be{chiral}
J=\sum_{\chi ,\,k} \chi :c_{\chi,\, k}^{\dag}\,c_{\chi,\, k}:
\ee
as well as the total particle number $N$ are fully conserved by the Hamiltonian  \pref{fullHam}.

%%%%%%%%%%
\section{Renormalized Luttinger parameters}
%%%%%%%%%%
The Hamiltonian \pref{fullHam} can be diagonalized by use of standard bosonization technique. Here we only briefly discuss the application of the method to the present case and refer to \cite{Haldane81} for more details. For given $N$ and $J$ all exitations of the system can be generated by the bosonic creation and annihilation operators defined by the chiral density operators,
\be{crean}
a_q=\sqrt{{2\pi}\over{|q|L}}\sum_\chi\theta(\chi\,q)\rho_{\chi,q}\;,\quad a_q^{\dag}=\sqrt{{2\pi}\over{|q|L}}\sum_\chi\theta(\chi\,q)\rho_{\chi,-q}\quad (q\neq 0)
\ee
With the Hamiltonian \pref{fullHam} separated in a one-body and a two-body part,
\be{1-2body}
H=H_{\rm 1part}+H_{\rm 2part}
\ee
the two-body part can readily be reformulated in terms of the bosonic operators if the $q=0$ component is separated from the $q\neq 0$ part and expressed in terms of the chiral particle number operators,
\be{Nchi}
N_{\chi}=\sum_k:c_{\chi,\,k}^{\dag}\,c_{\chi,\,k}:
\ee
The result is 
\be{2part2}
H_{\rm 2part}&=&\frac{\bar V_0}{4L}\sum_{\chi}( N_{\chi}^2+ e^{-4b^2k_F^2}N_{\chi}N_{-\chi})\nn
&+&\frac{1}{8\pi}\sum_{q\neq 0}|q|\left[(V_{1}(q)(a_q^{\dag}a_q+a_q a_{q}^{\dag})+V_{2}(q)(a_q^{\dag}a_{-q}^{\dag}+a_q a_{-q})\right] 
\ee
The one-body part of the Hamiltonian, with a linear momentum dependence, can also be re-expressed in terms of bosonic variables, although in a less obvious way. We only refer to the standard result,
\be{1part3}
H_{1part}=\bar v_F\frac{\pi}{L}\sum_{\chi} N_{\chi}^2+\bar v_F\sum_{q\neq 0}|q|a_q^{\dag} a_q
\ee

For the full Hamiltonian, with $N_{\chi}$ expressed in terms of $N$ and $J$, the bosonic form is
\be{fullHam2}
H&=&{{\pi\hbar}\over{2L}}(v_N(N-N_0)^2+v_JJ^2)
-{\hbar\over 2}\sum_{q\neq 0} \bar v_F|q|
\nn
&+&{\hbar\over 2}\sum_{q\neq 0}|q|\left[
(\bar v_F+\frac{V_{1}(q)}{4\pi\hbar})
(a_q^{\dag}a_q+a_q a_{q}^{\dag})+\frac{V_{2}(q)}{4\pi\hbar}(a_q^{\dag}a_{-q}^{\dag}+a_q a_{-q})\right] \nn
\ee
where the new velocity parameters $v_N$ and $v_J$ are given by
\be{velocities}
v_N&=&\bar v_F+ \frac{\bar V_0}{4\pi\hbar}(1+e^{-4b^2k_F^2})=v_F+\frac{\bar V_0}{2\pi\hbar}e^{-4b^2k_F^2}\nn
v_J&=&\bar v_F+ \frac{\bar V_0}{4\pi\hbar}(1-e^{-4b^2k_F^2})=v_F
\ee
One should here note the interesting point that the velocity parameter $v_J$ is not renormalized by the interaction. Thus, the contribution from the $q=0$ component of the interaction between the low energy electrons in \pref{fullHam} is exactly cancelled by the contribution from the effective potential created by the electrons in the Fermi sea. 

For $v_N$ there is a renormalization effect as long as the factor $\exp(-4b^2k_F^2)$ is non-negligible. This condition can be linked to the range of the interaction potential in 2D. Thus, if $\Lambda\approx 1$, with $\Lambda$ as the scale factor in \pref{rescaled}, then
\be{expfac}
e^{-4b^2k_F^2}\approx e^{-\ga^2 W^2}
\ee
with $1/\ga$ as the range of the 2D potential and $W$ as the width of the quantum Hall channel. This means that for a sufficiently short range potential, $1/\ga<<W$, with no communication between the two edges, there is no renormalization effect. In that case the interaction $V_2(q)$ becomes  negligible for all $q$ due to the relation
\be{V1-2} 
V_2(q)=e^{-\ga^2 W^2} V_1(q)
\ee
In the other limit, with $1/\ga>>W$, the interaction in 1D is changed to a local interaction, and there is no difference between the two interactions, $V_1(q)=V_2(q)$.
Therefore, the intermediate case is in a sense the most interesting one, with the range of the potential comparable to the distance between the two edges. In that case both $V_1$ and $V_2$ are different from zero, and they are not equal due to the interaction between the two edges.

The Hamiltonian \pref{fullHam2} is brought into the diagonal, free-field form, by a Bogoliubov transformation. This transforms the annihilation operators as
\be{Bog}
a_q=S^{\dag}b_qS=\cosh\gz_q\; b_q+\sinh\gz_q\;b_{-q}^{\dag}
\ee
with
\be{S}
S=\exp[\half\sum_{q\neq 0} \gz_q\,(a_q^{\dag}a_{-q}^{\dag}-a_qa_{-q})]
\ee
and $\gz_q$ defined by
\be{tanh}
\tanh(2\gz_q)=-\frac{V_{2}(q)}{V_{1}(q)+4\pi\hbar \bar v_F}=-e^{-\alpha^2W^2}\frac{\bar V_0 \exp(-\frac{q^2}{4\alpha^2})}{\bar V_0 \exp(-\frac{q^2}{4\alpha^2})+4\pi \hbar \bar v_F}
\ee
Expressed in terms of the transformed bosonic operators the Hamiltonian is
\be{transHam}
H=E_0+\hbar \sum_{q\neq 0}\omega_qb_q^{\dag}b_q+{{\pi\hbar}\over{2L}}(v_N(N-N_0)^2+v_JJ^2)
\ee
with
\be{E0}
E_0={\hbar\over 2}\sum_{q\neq 0}(\omega_q-\bar v_F|q|)
\ee
and
\be{omega}
\omega_q=\frac{1}{4\pi\hbar}\sqrt{(V_{1}(q)+4\pi \hbar \bar v_F)^2-V_{2}(q)^2}\;|q|
\ee

An important parameter of the Luttinger liquid theory, is the one that specifies the asymptotic behavior of the electron correlation function. Thus for large distance $\xi$, the correlation function falls off with distance as $1/ \xi^{\gamma}$. The parameter $\gamma$ depends on the interaction and can be determined by expressing the electron operators in terms of the bosonic variables. We quote the general result and apply it to the present case
\be{gamma}
\gamma&=&\sqrt{\frac{(V_{1}(0)+4\pi \hbar \bar v_F)^2}{(V_{1}(0)+4\pi\hbar \bar v_F)^2-V_{2}(0)^2}} \nn
&=&\sqrt{\frac{(\bar V_0+4\pi \hbar \bar v_F)^2}{(\bar V_0+4\pi\hbar \bar v_F)^2-\exp(-2\ga^2 W^2)\bar V_0^2}} \nn
\ee
One notes that the deviation of $\gamma$ relative to the non-interaction value $1$ depends on the interaction between particles of opposite chiralities, \ie between electrons of the two edges of the quantum Hall system. If we re-write the expression in terms of the unnormalized Fermi velocity $v_F$ and introduce the relative interaction strength,
\be{relstrength}
\Delta=\frac{V_2(0)}{4\pi \hbar v_F}=\frac{\exp(-\ga^2 W^2)\bar V_0}{4\pi \hbar v_F}
\ee
the expression for $\gamma$ is
\be{gamma2}
\gamma=\frac{1+\Delta}{\sqrt{1+2\Delta}}
\ee
For weak interaction between the two edges, with $\Delta<<1$, this gives to leading order,
\be{gamma3}
\gamma\approx 1+\half\Delta^2
\ee

For more details concerning the correlation functions we refer to \cite{Horsdal}. We there in particular examine how renormalization effects, due to interaction, modify the 2D correlation function when we map the system back from 1D to the description of the quantum Hall state.

%%%%%%%%%%
\section{Current renormalization and the Hall conductivity}
%%%%%%%%%%
The electron interaction will modify the expression for the current density, both in the 1D and 2D description. This follows from current conservation, since the time derivative of the charge density will be modified by the interaction. Expressed in terms of the Fourier components this relation is (in the low energy approximation)
\be{currentcons}
qj_q=\frac{1}{\hbar} \sum_{\chi}\com{H}{\rho_{\chi,\,q}}
\ee 
where the commutator can be determined from the bosonized form of the operators. For $q\neq 0$ the result is
\be{jq}
j_q=\sum_{\chi}\chi  (\bar v_F+\frac{V_1(q)}{4\pi\hbar}-\frac{V_2(q)}{4\pi\hbar})\rho_{\chi,\,q}
\quad (q\neq 0)
\ee
The $q=0$ component is not determined by the current conservation condition, which is obvious from eq. \pref{currentcons}. However, if one simply assumes that the expression  \pref{jq} can be extended to $q=0$, this gives for the integrated current  \cite{Haldane81},
\be{jJ}
j_0=v_J J
\ee
where $v_J$ is the same velocity parameter as in the expression for the $J$-dependent contribution to the energy. As already discussed, $v_J$ is changed by the interactions, relative to the modified Fermi velocity $\bar v_F$. However, relative to the original Fermi velocity $v_F$, which is determined by the confining potential alone, there is no renormalization effect, $v_J=v_F$.

The question about a possible renormalization of the current is an interesting one with relation to the quantum Hall effect in the 2D description. With a long range interaction that connects the two edges of the quantum Hall sample, the effect of the interaction on the 2D electron system is not completely obvious, but the non-renormalization of $v_J$ in 1D is consistent with the conclusion that the Hall conductivity is left unchanged by a (weak) long range interaction. We will discuss this point, and assume for simplicity that  $\omega<<\omega_c$, so we can neglect the difference between the true and the effective magnetic field for the electrons in the harmonic oscillator potential.

Consider then the 2D system in a situation where a weak electric field is applied in the $y$ direction. This modifies the external potential as
\be{newpot}
\half m \omega^2 y^2 &\to& \half m \omega^2 y^2-eEy\nn
&=& \half m \omega^2 (y-\frac{eE}{m\omega^2})^2-\frac{(eE)^2}{2m\omega^2}
\ee
with $E$ as the electric field strength. Thus, the only change is a shift of the minimum of the potential in the $y$ direction and the addition of a small constant term. The new ground state is therefore the same, up to a shift in the $y$-direction, as without the electric field. 
The shift in the $y$ direction means that it has the non-vanishing chiral charge relative to the original ground state,
\be{chiral2}
J=2\frac{eE}{m\omega^2}L \rho_0
\ee
where $\rho_0$ is the bulk particle density in the quantum Hall state. The strength of the electric field may be adjusted to make $J$ an even integer, in which case the new ground state is an energy eigenstate also of the old Hamiltonian (with $E=0$). It is not the true (global) ground state, but rather the ground state in a $J\neq 0$ sector.

An explicit transformation from the $E\neq 0$ system to the $E=0$ system may be introduced by a change from a fixed frame to a frame that moves with velocity $v$ in the x-dirextion. Thus, if the velocity is chosen as $v=E/B$ the transformed electric field vanishes, while the $B$ field is modified by terms that are negligibly small for sufficiently small $E$ and $\omega$. 

 Let us consider the linear current density $j(x)=\int j_x(x,y) dy$, with $j_x(x,y)$ as the x-component of the planar current density. In the ground state (for $E\neq 0$) we obviously have $j(x)=0$. However, in the moving frame, with $E=0$, the current is non-vanishing.
The current density in this frame has the constant value
\be{j}
j= v N/L =  \frac{E}{B} \rho_0 W
\ee
where $N$ is the particle number, and the width of the quantum Hall channel here is defined as $W=N/(\rho_0 L)$. Therefore, in this frame both $J$ and $j$ are non-vanishing, with the relation
\be{jJrel}
j=\frac{m\omega^2}{eB}W/L \;J 
\ee
If we further introduce the Fermi velocity $v_F=(\hbar k_F/m)(\omega/ \omega_c)^2$ and  and assume the width $W$ to be the same as in the non-interacting case, $W=2l_B^2 k_F$, then the relation simplifies to
\be{jJrel2}
j= v_F J/L
\ee
This relation for the 2D current translates directly to eq. \pref{jJ} in the 1D case, with $j_0 = \int j(\xi) d\xi = j L$ and $v_J$ given by the unrenormalized value $v_F$. (There is a small non-locality in the mapping between the 1D and 2D current densities, but this is of no importance for the integrated current.)

Thus, the relation between $j$ and $J$ given by \pref{jJrel} is consistent with the result $v_J=v_F$, provided the width $W$ and therefore the particle density is left unchanged by the interactions. This further means that the Hall  conductivity is not modified. Thus from \pref{j} follows that the Hall conductivity is
\be{Hall}
\sigma= \frac{j}{EW}=\frac{e^2}{2\pi\hbar}
\ee
which is the quantized Hall conductivity at integer filling, without any correction due to interactions. 

As a final comment on the question of the renormalization of the current due to interactions, we return to the 1D result and write the current density in a standard many-body form. For the integrated current this is
\be{j0}
j_0=\int d\xi \sum_i v_i\,\delta(\xi_i-\xi)=\sum_i v_i = \frac{P}{M}
\ee
where the summation is over all the particles in the system, and $P$ denotes the total momentum. This shows that for interactions that respect conservation of total momentum, one should not expect renormalization of $j_0$ due to interactions.  A related, general argument is given in ref.\cite{Haldane81b}, that $v_J$ is not renormalized by interactions in a 1D many-particle system with Galilean invariance. In the present case, the system of 2D electrons in the magnetic field is not Galilean invariant, but even so the corresponding 1D system, with interactions \pref{kspace}, is invariant under both translations and Galilean transformations.

%%%%%%%%%%
\section{Conclusions}
%%%%%%%%%%
The model discussed in this paper gives the possibility of making an explicit mapping of a 2D quantum Hall system to  a 1D electron system, which in the low energy approximation takes the form of a Luttinger liquid. We have examined this mapping with the focus on the low-energy description of the particle interactions, and on how the  physical parameters of the quantum Hall fluid determine the Luttinger parameters. 

The model is simple and idealized, in order to make explicit evaluations possible. Thus, we assume an harmonic confining potential for the 2D electrons, with a weak electron interaction. However, these simplifications do not seem very important for the main conclusions. With a stronger interaction between the electrons, the confining  potential may simply include the effects of the  Fermi sea, and the conclusions concerning renormalization effects for the Luttinger parameters do not seem to depend strongly on the harmonic oscillator form of the background potential.

We have assumed that the particle interaction is sufficiently long range to introduce interactions between electrons at opposite edges of the quantum Hall fluid. In the low energy approximation that means that there are non-vanishing interactions between particles of opposite chirality as well as between particles of equal chirality.  Since these two types of interactions are different, that may suggest that the two low-energy parameters, $v_N$ and $v_J$, are both modified by the interactions. However, that is not the case, as the detailed analysis shows. Even if the interaction depends on two parameters, the strength and the range, only a single combination of these appears in the low energy description in such a way that $v_J$ unchanged. In our analysis that can be explained as a cancelation effect when interactions between the low energy particles and particles in the Fermi sea are included in addition to interactions between the low energy particles.

The non-renormalization of the current parameter $v_J$ we have further shown to be consistent with general arguments applied to the model we study. On one hand we link that to robustness of the Hall conductivity. Thus, even if the interactions are long range and extend across the the electron system, the fact that $v_J$ is not modified by interactions is consistent with the assumption that that the Hall conductivity is unchanged. On the other hand, the fact that momentum is conserved indicates that the integrated current, and therefore $v_J$ will not be modified by the interactions.

There are other effects of the interaction, for example for the density profile at the edges of the system and for the asymptotic behaviour of the electron correlation function. We refer to the subsequent paper \cite{Horsdal}, for a discussion of these questions. In this paper we evaluate the correlation function of the 2D interacting system by use of the 1D Luttinger liquid description.\\

\noindent
{\bf Acknowledgement}\\
We thank T.H. Hansson for helpful discussions.
This work has been supported by NordForsk.

%%%%%%%%%%

%%%%%%%%%
%%%%%%%%%%

\begin{thebibliography}{99}
%%%%%%%%%%

\bibitem{Wen92}
X.G. Wen, {\em Theory of the edge states in fractional quantum Hall effects}, Int. J. Mod. Phys. B {\bf 6}, 1711 (1992)

\bibitem{Goldman01}
V.J. Goldman and E.V. Tsiper, {\em Dependence of the fractional quantum Hall edge critical exponent on the range of the interaction}, Phys. Rev. Lett. {\bf 86}, 5841 (2001).

\bibitem{Jain01}
S.S. Mandal and J.K. Jain, {\em How universal is the fractional-quantum-Hall edge Luttinger liquid?}, Solid State Commun. {\bf 118}, 503 (2001)

\bibitem{WanEvers04}
X. Wan, F. Evers and E.H. Rezayi, {\em Universality of the edge tunneling exponent of fractional quantum Hall liquids}, Phys. Rev. Lett. {\bf 94}, 166804 (2005).

\bibitem{Haldane81}
F.D.M. Haldane, {\em Luttinger liquid theory of one dimensional quantum fluids: I. Properties of the Luttinger model and their extension to the general 1d interacting spinless Fermi gas}, J. Phys. C {\bf 14}, 2585 (1981).

\bibitem{Horsdal}
M. Horsdal and J.M. Leinaas, {\em Explicit mapping between a 2D quantum Hall system and a 1D Luttinger liquid, II. Correlation functions}, in preparation.

\bibitem{Haldane81b}
F.D.M. Haldane, {\em Effective harmonic fluid approach to low energy properties to one-dimensional quantum fluids}, Phys. Rev. Lett. {\bf 47}, 1840 (1981).

\end{thebibliography}
\end{document}